\documentclass[aps,twocolumn,secnumarabic,balancelastpage,amsmath,amssymb,nofootinbib,hyperref=pdftex]{revtex4}

\usepackage{chapterbib}    
\usepackage{color}         
\usepackage{graphics}      
\usepackage[pdftex]{graphicx}      
\usepackage{longtable}     
\usepackage{epsf}          
\usepackage{bm}            
\usepackage{verbatim}			
\usepackage[colorlinks=true]{hyperref}  
\usepackage[english]{babel}
\usepackage{graphicx}

\usepackage{tikz} 
\tikzset{inner sep=0pt, 
  root/.style={circle,draw,minimum size=7pt,thick}, 
  short root/.style={circle,fill,minimum size=7pt}, 
  double/.style={double distance=2pt,thick}, 
  doublearrow/.style={postaction={decorate}, 
  decoration={markings,mark=at position .7
  with {\arrow{angle 60}}},double distance=3pt,thick}
  \setlength{\unitlength}{1mm}
}
\usetikzlibrary{calc}
\usepackage{amssymb}
\usetikzlibrary{decorations.markings}
\usepackage{simplewick}

\newcommand{\diagramb}[1]{
\begin{center}
\begin{tikzpicture}[scale=1]

 \draw[decoration={markings, mark=at position 0.125 with {\arrow{>}}}, postaction={decorate}] (0,0) circle (0.5);
  \draw[decoration={markings, mark=at position 0.125 with {\arrow{<}}}, postaction={decorate}] (0,0) circle (1);
 \fill (0,0) circle (0.03);
 \fill (0.55,0.55) circle (0.03);
\node [below right] at (-0.05,0.05) {\scalebox{0.7}{$z_0$}};
\node [below right] at (0.50,0.6) {\scalebox{0.7}{$z$}};
\node [below right] at (0.24,-0.24) {\scalebox{0.7}{$\mathcal{C}_1$}};
\node [below right] at (0.5,-0.7) {\scalebox{0.7}{$\mathcal{C}_2$}};

\draw[|->,semithick] (1.5,0) -- (2.5,0);

 \draw[decoration={markings, mark=at position 0.125 with {\arrow{<}}}, postaction={decorate}] (3.85,0.55) circle (0.4);
 \fill (3.3,0) circle (.03);
 \fill (3.85,0.55) circle (.03);
\node [below right] at (3.25,0.05) {\scalebox{0.7}{$z_0$}};
\node [below right] at (3.80,0.6) {\scalebox{0.7}{$z$}};
\node [below right] at (4.0,0.3) {\scalebox{0.7}{$\mathcal{C}_1  \cup \mathcal{C}_2$}};

\end{tikzpicture}
\end{center}
}

\newcommand{\diagrama}[1]{
\begin{tikzpicture}[scale=.7]
	\draw (-0.2,3.5) node {$M=S^1 \times \mathbb{R}$};
	\draw[dashed,color=gray] (2.9,0) arc (-90:90:-0.5 and 1.5);
	\draw[dashed,color=gray] (1.4,0) arc (-90:90:-0.5 and 1.5);
	\draw[->, semithick, color=gray] (-0.5,0) arc (-90:90:-0.5 and 1.5);
	\draw[semithick] (0,0) -- (4,0);
	\draw[semithick] (0,3) -- (4,3);
	\draw[semithick] (0,0) arc (270:90:0.5 and 1.5);
	\draw[semithick] (4,1.5) ellipse (0.5 and 1.5);
	\draw (0.7,1.5) node {$t_1$};
	\draw (2.2,1.5) node {$t_2$};
	\draw (-1.25,1.5) node {$x$};
	\draw[->,semithick] (0.5,-0.5) -- (3.5,-0.5);
	\draw (2,-1) node {$t$};
	
	\draw[|->,thick] (5.0,1.5) -- (6,1.5);
	
	\draw[dashed] (9, 1.5) circle (0.8);
	\draw[dashed] (9, 1.5) circle (1.8);
	\draw [<-,semithick,domain=0:90] plot ({9 - 2.2*cos(\x)}, {1.5-2.2*sin(\x)});
	\draw (6.9,0.2) node {$x$};
	\draw (9.6,2.4) node {$t_1$};
	\draw (10.4,3) node {$t_2$};
	\draw (7.3,3.5) node {$M= \mathbb{C} $};
	\node at (9,1.5) [circle, inner sep=1pt, fill=black] {};
\end{tikzpicture}
}

\newcommand{\diagram}[1]{

\begin{center}
  \begin{tikzpicture}[scale=.4]
    \draw (-1,1.5) node[anchor=east]  {$\hat{A}_n$};
    \foreach \x in {0,...,5}
    \draw[xshift=\x cm,thick] (\x cm,0) circle (.3cm);
    \draw[dotted,thick] (0.3 cm,0) -- +(1.4 cm,0);
    \foreach \y in {1.15,...,4.15}
    \draw[xshift=\y cm,thick] (\y cm,0) -- +(1.4 cm,0);
        \draw[thick] (5 cm,3 cm) circle (3 mm);
    \draw[thick] (0.2 cm, 3mm) -- +(4.62, 2.52 cm);
    \draw[thick] (10 cm, 3mm) -- +(-4.8, 2.52 cm);
  \end{tikzpicture}
\end{center}

\begin{center}
  \begin{tikzpicture}[scale=.4]
    \draw (-1,0) node[anchor=east]  {$\hat{B}_n$};
    \draw[thick] (0 cm, 0.8 cm) circle (3 mm);
    \draw[thick] (0 cm, -0.8 cm) circle (3 mm);
    \foreach \x in {1,...,5}
    \draw[xshift=\x cm,thick] (\x cm,0) circle (.3cm);
    \draw[dotted,thick] (2.3 cm,0) -- +(1.4 cm,0);
    \foreach \y in {2.15,...,3.15}
    \draw[xshift=\y cm,thick] (\y cm,0) -- +(1.4 cm,0);
    \draw[thick] (8.3 cm, .1 cm) -- +(1.4 cm,0);
    \draw[thick] (8.3 cm, -.1 cm) -- +(1.4 cm,0);
    \draw[xshift=1.5 cm,thick] (30: 3 mm) -- (-30: -14 mm);
    \draw[xshift=1.5 cm,thick] (-30: 3 mm) -- (30: -14 mm);

  \end{tikzpicture}
\end{center}

\begin{center}
  \begin{tikzpicture}[scale=.4]
    \draw (-1,0) node[anchor=east]  {$\hat{C}_n$};
    \draw[thick] (0.3 cm, .1 cm) -- +(1.4 cm,0);
    \draw[thick] (0.3 cm, -.1 cm) -- +(1.4 cm,0);
    \foreach \x in {0,...,4}
    \draw[xshift=\x cm,thick] (\x cm,0) circle (.3cm);
    \draw[xshift=5 cm,thick] (5 cm, 0) circle (.3 cm);
    \draw[dotted,thick] (2.3 cm,0) -- +(1.4 cm,0);
    \foreach \y in {2.15,...,3.15}
    \draw[xshift=\y cm,thick] (\y cm,0) -- +(1.4 cm,0);
    \draw[thick] (8.3 cm, .1 cm) -- +(1.4 cm,0);
    \draw[thick] (8.3 cm, -.1 cm) -- +(1.4 cm,0);
  \end{tikzpicture}
\end{center}

\begin{center}
  \begin{tikzpicture}[scale=.4]
    \draw (-1,0) node[anchor=east]  {$\hat{D}_n$};
        \draw[xshift=1.5 cm,thick] (30: 3 mm) -- (-30: -14 mm);
    \draw[xshift=1.5 cm,thick] (-30: 3 mm) -- (30: -14 mm);
    \draw[thick] (0 cm, 0.8 cm) circle (3 mm);
    \draw[thick] (0 cm, -0.8 cm) circle (3 mm);
    \foreach \x in {1,...,4}
    \draw[xshift=\x cm,thick] (\x cm,0) circle (.3cm);
    \draw[xshift=8 cm,thick] (30: 17 mm) circle (.3cm);
    \draw[xshift=8 cm,thick] (-30: 17 mm) circle (.3cm);
    \draw[dotted,thick] (2.3 cm,0) -- +(1.4 cm,0);
    \foreach \y in {2.15,...,3.15}
    \draw[xshift=\y cm,thick] (\y cm,0) -- +(1.4 cm,0);
    \draw[xshift=8 cm,thick] (30: 3 mm) -- (30: 14 mm);
    \draw[xshift=8 cm,thick] (-30: 3 mm) -- (-30: 14 mm);
  \end{tikzpicture}
\end{center}

}

\begin{document}
\title{Group Theoretical Approach to the Construction of Conformal Field Theories}
\author         {Benjamin Horowitz}
\email          {benjamin.a.horowitz@yale.edu}
\homepage{http://pantheon.yale.edu/~bah8/Ben_H.html}
\date{\today}
\affiliation{Yale University, Department of Physics}

\begin{abstract}
A conformal field theory (CFT) is a quantum field theory which is invariant under conformal transformations; a group action that preserve angles but not necessarily lengths. There are two traditional approaches to the construction of CFTs: analyzing a statistical system near a critical point as a euclidean field theory, and in holographic duality within the context of string theory. This pedagogical paper presents a  construction of CFTs using purely group theoretic techniques. Starting with the basic definition of a Lie algebra and quantum field theory, we generalize to affine Lie algebras and form a  energy momentum tensor via the Sugawara construction. 
\end{abstract}

\maketitle


\section{Introduction}

The purpose of studying two-dimensional conformal field theories is many fold. First, all two dimensional CFTs are solvable non-perturbatively, and are useful in the study of physical field theories in non-perturbative regimes (i.e. strongly coupled theories). CFTs also have found widespread applications in statistical mechanics and condensed matter systems, where second order phase transitions (like those at the paramagnetic-ferromagnetic boundary) exhibit large scale correlations and, according to Polyakov's seminal paper, \cite{belavin1984infinite} exhibit symmetries beyond simple scalings: the conformal group. Polyakov along with Belavin and Zamolodchikov, showed how an infinite-dimensional field theory problem could be reduced to a finite problem, by the presence of an infinite-dimensional symmetry with an associated Virasoro algebra. This algebra can be viewed as the central extension of polynomial vector fields on the circle and has found widescale applications to the study of conformal field theories and String Theory. 

	This paper discusses the construction of these algebras by first discussing the construction of Kac-Moody Algebras and the associated operator product of their realizations. Then the energy momentum tensor, a critical component of any theory wishing to have physical significance, arises from the Sugawara construction. Finally, these tools are applied in the context of the conformal group, and its associated two-dimensional conformal field theories.
\section{Simple and Affine Lie Algebras}

	Simple Lie algebras emerge naturally in physics to describe important relationships and symmetries between physical quantities. Commutator relationships between elements of the angular momentum operator form a Lie algebra, and irreducible representations of the Poincare group can be used to describe various quantum states of elementary particles. A general finite dimensional simple Lie algebra can be completely generated by $3k$ generators $ \{E^i_{\pm}, H^i | i = 1, . . . , k\}$ which satisfy the following \emph{Chevalley-Serre relations};
\begin{equation}
\begin{split}
[H^i,H^j]=0, \\ [H^i,E^i_{\pm}] = \pm A^{ji}E^j_{\pm}, \\  [E^i_+ , E^j_-]= \delta^{ij}H^j, \\ (ad_{E^i_{\pm}})^{1-A^{ji}}E^j_{\pm} = 0,\:\: i\neq j.
\end{split}
\label{eq:C-S}
\end{equation}
From this description, we see that every Lie algebra can be uniquely determined from the components of the $k \times k$ \emph{Cartan matrix}, $A^{ji}$, which has the following constraints
\begin{equation}
\begin{split}
A^{ii} = 2, \\ A^{ij} \leq 0, \:\: i \neq j, \\ A^{ij} =0 \Longleftrightarrow A^{ji} = 0, \\ A^{ij} \in \mathbb{Z},
\label{eq:C-M}
\end{split}
\end{equation}
and
\begin{equation}
det A > 0.
\label{eq:C-Ms}
\end{equation}
This last constraint is weakened in the case of \emph{Kac-Moody algebras}, of which \emph{affine Lie algebras} are a sub-class. Abstractly, a Kac-Moody algebra is the set of all smooth mappings of a manifold $\mathcal{M}$ to a finite-dimensional Lie algebra. In our case, we wish to look at \emph{loop algebras} where $\mathcal{M} = S^1$. The resulting algebra has the same properties detailed in (2), but condition (3) is replaced by 
\begin{equation}
det A_{\{i\}} > 0,  \:\: \forall i = 1,...,k ,
\label{eq:C-M}
\end{equation}
where $A_{\{i\}}$ is the $i$th principal minor of $A$ (the $i$th column and row are removed from the matrix). Any algebra which satisfies conditions (1), (2), and (4) is considered an affine Lie algebra. Like simple Lie algebras, one can construct Dynkin diagrams based off the root system the commutator relations create. With the new constraint, we have a new set of diagrams shown in figure 1.
\begin{figure}[hi]
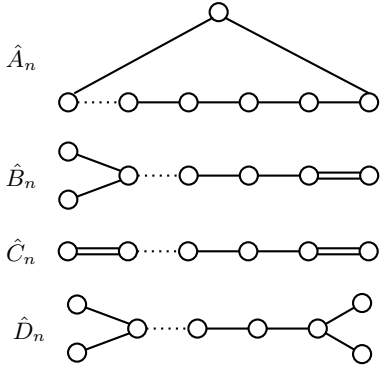

\diagram
\caption{\emph{Figure 1}: Dynkin diagrams of the non-exceptional affine Lie algebras.}
\label{fig:dynk}
\end{figure}

Affine Lie algebras are more ``complicated" than simple Lie algebras in that they admit non-trivial centers; in particular they have a \emph{central element}, $K \in \mathcal{Z}(g)$, 
\begin{equation}
\begin{split}
[K,E^i_\pm]=0, \\ [K,H^i] = 0, 
\end{split}
\label{eq:central_id}
\end{equation}
defined as 
\begin{equation}
K = \beta \sum_{i=0}^{k} a_i^{\vee} H^i, \: \: \beta \in \mathbb{R} ,
\label{eq:central}
\end{equation}
where $a_i^{\vee}$ is the dual Coxeter element defined in terms of the Cartan matrix by
\begin{equation}
\sum_{j=0}^k A^{ij} a_j^{\vee} = 0.
\label{eq:coxeter}
\end{equation}
The simple Lie bracket relationship of a generic element, $[T^a,T^b]=f^{ab}_cT^c$, can be generalized to 
\begin{equation}
[T^a,T^b]=f^{ab}_cT^c +f^{ab}_i K^i.
\label{eq:coxeter}
\end{equation}
The structure constant of this new term, $f^{ab}_i$, is critical as the dimensionality of the space of solutions of $f^{ab}_i$ is equal to the number of allowed central extensions of the algebra, $g$. These extensions are well visualized by going back to the definition of a Kac-Moody algebra as the space of analytic maps from $S^1$ to $g$. Given the Lie algebra algebra basis for $g$, $\{T^a \: | \: a = 1, ..., k\}$, we can construct a basis for the vector space generated by those maps by 
\begin{equation}
\begin{split}
\{T^a_n = T^a \otimes z^n \: | \: a = 1, ..., k, \: n \in \mathbb{Z}, \: |z|=1 \}.
\end{split}
\label{eq:basis}
\end{equation}
It is trivial to show that the the commutator relations generalize correspondingly to
\begin{equation}
\begin{split}
[T^a_m,T^b_n]=f^{ab}_cT^c_{m+n} +(f^{ab}_i)_{mn} K^i.
\end{split}
\label{eq:basis}
\end{equation}
One can constrain the new structure constant term, $(f^{ab}_i)_{mn}$ a number of ways. In the limiting case of $n=0$, we should recover the original simple algebra since the $z^m$ term can be factored out the commutator. We also know that the basis elements are invariant in the adjoint representation of the original Lie algebra $g_0$, so the indices $a,b$ are invariant under this algebra. The only rank two tensor with this property is the metric, or Killing form, of the algebra defined in terms of the adjoint representation as
\begin{equation}
\kappa^{xy} = tr(ad_x \circ ad_y),
\label{eq:killing}
\end{equation}
where $ad_x$ represent the matrix of the $x$ element in the adjoint representation. Since the Killing form is the only such element, we only have one central extension and the index $i$ is equal to $1$. These conditions combine to restrict the structure constant to a much more manageable form
\begin{equation}
(f^{ab}_i)_{mn} = m \delta_{m+n,0} \kappa^{ab},
\label{eq:constant}
\end{equation}
and produce a new commutator relationship,
\begin{equation}
[T^a_m,T^b_n]=f^{ab}_cT^c_{m+n} + m \delta_{m+n,0} \kappa^{ab} K.
\label{eq:constant}
\end{equation}

It is important to note that this reduction is only possible because a loop algebra has only one central extension. These algebras have great physical significance since they can be used to describe and categorize integrable systems, but more general Kac-Moody algebras can be constructed with more complicated manifolds, as with multi-loop algebras. \cite{senesi}


\section{Operator Products}

	This definition of the commutator relation, however, lacks a degree of physical significance and is generally unwieldy. We wish to simplify our calculations by introducing a \emph{affine current operator} over some field, $J^a(z)$, which is the generating function of the $T_m^a$ algebra elements defined as
	
\begin{equation}
J^a(z) = \sum_{n \in \mathbb{Z}} T^a_n z^{-n-1}.
\label{eq:C-Mx}
\end{equation}
	
	The field $z$ can be given physical significance by viewing it as a parameter of the unit circle $S^1 \subset \mathbb{C}$ on which we define some conformally invariant theory. With this interpretation, one can identify the expansion given in Eq.~\ref{eq:C-Mx} as a Laurent series obtained by expanding $J^a(z)$ around the origin, $z_0$, giving 
\begin{equation}
T^a_n = \frac{1}{2\pi i} \oint_{z_0} dz\: z^n J^a(z).
\label{eq:integral}
\end{equation}
	Using this definition we can construct the commutator relations of the basis elements in terms of the currents by
	
\begin{equation}
\begin{split}
[T^a_n,T^b_m] = \left(\frac{1}{2\pi i} \right)^2 \left[ \oint_{z_0} \! \! dz \oint_{z_0} \! \! dw -  \oint_{z_0} \! \! dw \oint_{z_0} \! \! dz \right] \\\ \times z^m w^n J^a(z)J^b(w),
\end{split}
\label{eq:basis_com}
\end{equation}
	
	and the commutators of the affine current fields by
	
	\begin{equation}
\begin{split}
[J^a(z),J^b(w)] = \sum_{m,n \in \mathbb{Z}} z^{-m-1}w^{-n-1} [T^a_m,T^b_n] \\ = \sum_{n \in \mathbb{Z}} \left( \frac{z}{w} \right)^{n+1} \! \! \! \! \! \sum _ {m+n\in \mathbb{Z}} z^{-(m+n) -1} [T^a_m, T^b_n].\end{split}
\label{eq:com_simp}
\end{equation}
	
	Notice that this sum, at first, doesn't look convergent as the term $(z/w)^{n+1}$ will explode as in either the case of $n+1$ going to infinity if $|z|>|w|$ or in the case $n+1$ going to negative infinity for $|z|<|w|$. 
		
\begin{figure}[hi]
\includegraphics[width=8cm]{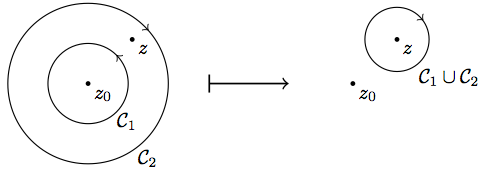}
\emph{Figure 2}: Construction of path integrals when looking at the commutator of two current operators.
\end{figure}

	One would like to re-write expression~\ref{eq:basis_com} in a form involving integration over current and field terms such that we get a linear combination of $T^i_p$ terms, as in Eq.~\ref{eq:constant}. We can accomplish this by looking at the integrals as over different circles containing $z_0$, such as $\mathcal{C}_1$ and $\mathcal{C}_2$ in figure 3; the first pair is to be preformed whenever $|z| > |w|$ and the second when $|z| < |w|$. We can define the \emph{radially ordered product} of two fields $a(z)$ and $b(z)$ as
	
\begin{equation}
\mathcal{R}[a(z)b(w)] = \begin{cases}
       a(z)b(w) & : |z| > |w| \\
       b(w)a(z) & : |z| < |w|
     \end{cases}
\label{eq:rop}
\end{equation}

One can rewrite equation Eq.~\ref{eq:basis_com} using this operator as

\begin{equation}
\begin{split}
[T^a_n,T^b_m] = \left(\frac{1}{2\pi i} \right)^2  \oint_{z_0} \! \! dz [ \oint_{\mathcal{C}_1} \! \! dw \:  z^m w^n J^a(z)J^b(w)  \\  -  \oint_{\mathcal{C}_2} \! \! dw  \: z^m w^n J^a(z)J^b(w)] \\ = \left(\frac{1}{2\pi i} \right)^2  \oint_{z_0} \! \! dz \: \oint_{\mathcal{C}_1 \cup \mathcal{C}_2} \! \! dw \: z^m w^n \mathcal{R}(J^a(z)J^b(w)).
\end{split}
\label{eq:com_com}
\end{equation}
	
	The contour integral of the radial ordering of the current fields can be expressed in a simple form by noticing that the majority of terms of their series expansion are analytic functions, $\mathcal{O}((z-w)^0)$ and will go to zero after the integration:
	
\begin{equation}
\begin{split}
\mathcal{R}(J^a(z)J^b(w)) = (z-w)^{-2} \kappa^{ab} K - (z-w)^{-1}\\ \times f^{ab}_{c} J^{c}(w)  + \mathcal{O}((z-w)^0).
\end{split}
\label{eq:com_com}
\end{equation}	

We can define a new spatial coordinate via the exponential map of $z = exp(2\pi i x / L)$ and define a new current operator as
\begin{equation}
\tilde{J}^a(x) = \frac{z}{L} J^a(z).
\label{eq:tilde}
\end{equation}
The sum over $n$ that appears in Eq.~\ref{eq:com_simp} can be further simplified down by noticing that
\begin{equation}
\begin{split}
\sum_{n \in \mathbb{N}} e^{2\pi i n z / L} = L \: \delta(x),
\end{split}
\label{eq:exp}
\end{equation}	
so
\begin{equation}
\begin{split}
[\tilde{J}^a(x),\tilde{J}^b(y)] = \frac{1}{2\pi i} \kappa^{ab} K \delta ' (x-y) \\ + f^{ab}_c \tilde{J}^c(x) \delta(x -y).
\end{split}
\label{eq:exp}
\end{equation}	
Relationships of this form that arise in the commutation relations of local fields in quantum field theory are known as \emph{current algebras}. Currents in these theories transform as a vector with respect to the space-time Lorentz algebra. For example, in non-Abelian Yang-Mills theories, these sorts of relations arise in the charge density commutator,
\begin{equation}
\begin{split}
[\rho^a(\vec{x}),\rho^b(\vec{y})] =  f^{ab}_i \rho^i(x) \delta(\vec{x} -\vec{y}).
\end{split}
\label{eq:exp}
\end{equation}

	\section{Normal Ordering and Wick Theorem}
	Recall the \emph{Wick ordering}, or \emph{normal ordering}, of bosonic operators is defined by a re-arrangement of operators to place all creation operators, $\hat{b}^\dagger$, to the left of all annihilation operators, $\hat{b}$ in the product. These operators, when evaluated on the field at a given point in the space, represent excitations of the field (particles) and one can use them to calculate scattering amplitudes in a generic quantum field theory. This special ordering is designed to avoid an unpleasant situation of accidentally annihilating $|0\rangle$ when looking at an expectation value in relation to the vacuum state,
\begin{equation}
\begin{split}
:  \hat{b}^\dagger\hat{b} :  \: =\: \hat{b}^\dagger\hat{b} \\ : \hat{b}\hat{b}^\dagger: \: = \: \hat{b}^\dagger\hat{b}. 
\label{eq:NO}
\end{split}
\end{equation}
	We can define a conformal analogue in the context of current algebras by noticing we can expand the radial ordering product into its own Laurent series in terms of fields $P_n(z)$,
\begin{equation}
\mathcal{R}(a(z)b(w)) = \sum_{n = -n_0}^\infty (z-w)^n P_n(w),
\label{eq:R_E1}
\end{equation}	
	and select
\begin{equation}
P_0(w) = \: \: : a(w)b(w) : ,
\label{eq:R_A}
\end{equation}	
which is analogous to the definition~\ref{eq:NO} in the case of a free field theory, as in that case there exists an expansion of the field operators in terms of creation and annihilation operators, and the $P_0$ term will never contain singular terms so it defines an equivalent general regularization procedure. We can also define the contraction of two fields as

\begin{equation}
\contraction{}{a(z)}{}{b(w)} a(z)b(w) =  \sum_{n = -n_0}^{-1} (z-w)^n P_n(w),
\label{eq:R_con}
\end{equation}	

so in totality the radially ordering operator, Eq.~\ref{eq:rop} can be expressed as

\begin{equation}
\mathcal{R}(a(z)b(w)) = \contraction{}{a(z)}{}{b(w)} a(z)b(w) +  : a(z)b(w) : + \: \mathcal{O}(z-w).
\label{eq:R_Exp}
\end{equation}	

As in the case of the current operators, a closed form solution for the normal ordering product can be found via Cauchy's Integral Theorem as

\begin{equation}
: a(w)b(z) : \: \: = \frac{1}{2 \pi i} \oint_z (w-z)^{-1}\mathcal{R}(a(w)b(z))
\label{eq:cauchy}
\end{equation}	

and

\begin{equation}
: [a(z),b(z)] : \:\: = \sum_{n=1}^{n_0} \frac{(-1)^{n+1}}{n!} \partial^nP_{(-n)}(z),
\label{eq:cauchy}
\end{equation}	
or, in the specific case of a current field,
\begin{equation}
: [J^a(z),J^b(z)] : \:\: = f^{ab}_{c} \partial J^c(z).
\label{eq:current_c}
\end{equation}	

We can use these results to generate a CFT version for free fields of the Wick Theorem (a procedure used to reduce arbitrary products of operators into a manageable form)

\begin{equation}
\begin{split}
\contraction{}{a(y)}{:b(z)}{c(z))} a(y):b(z)c(z): \: \: = \frac{1}{2 \pi i} \oint _z dw (w-z)^{-1} \\ \times [ \mathcal{R}(\contraction{}{a(y)}{}{b())}a(y)b(w)c(z)- \mathcal{R}(\contraction{b(w)}{a(y)}{}{c())}b(w)a(y)c(z))].
\end{split}
\label{eq:wick}
\end{equation}	

\section{Radial Ordering and Relativistic Strings}

	In quantum field theories, one also has a time ordering product defined for bosonic operators as:
	\begin{equation}
\begin{split}
\mathcal{T}[A(x)B(y)] = \begin{cases}
       A(x)B(y) & : x_0 > y_0 \\
       B(y)A(x) & : x_0 < y_0
     \end{cases}
\label{eq:TO}
\end{split}
\end{equation}

	The radial operating operator is an equivalent construction, where instead of explicit time dependence, one has ordering via $|z|$. The space can be defined as a time parameter (the modulus), $ \mathbb{R} $, fibered trivially over the spatial coordinate, $S^1$.  Alternatively, one can construct the retract map from $\mathbb{C}\setminus \{0\}$ to $S^1$ by the mappings

\begin{equation}
\begin{split}
z = e^{t+ i x},\\ \overline{z} = e^{t - i x},
\label{eq:iso}
\end{split}
\end{equation}
 where $t$ is a time coordinate and $x$ is a spacial coordinate along the circle. In either case, we get a cylindrical topology, $S^1 \times \mathbb{R} $. This description is equivalent to the space found in the theory of closed relativistic strings since space of the time evolution of a string, the worldsheet, has cylindrical topology. In general for quantum field theories, these mapping are difficult to construct as preserving both conformal invariance and causality is problematic as it must be defined on a $S^{d-1} \times \mathbb{R}$ covering of $d$-dimensional Minkowski space. This leads to the need to compactify $\mathbb{R}^{d-1,1}$ space.
	
	\begin{figure}[hii]
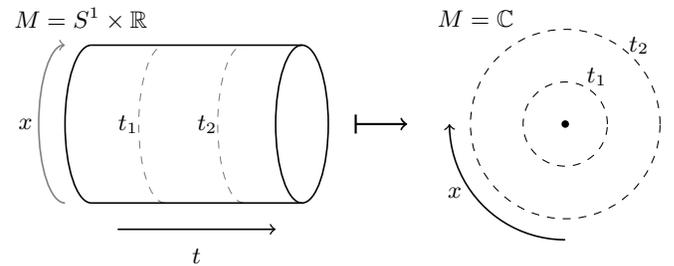

\diagrama
\caption{\emph{Figure 2}: Equivalence of radial ordering operator and time ordering operator.}
\end{figure}

\section{Sugawara Construction}

	A critical part of any physically relevant field theory is the existence of a spin two operator, the energy-momentum tensor. One approach to creating such a tensor is the \emph{Sugawara} construction given by

\begin{equation}
T(z)= \lambda \: \kappa^{ab}:J^a(z)J^b(z):.
\label{eq:sugawara}
\end{equation}

Note that this field also functions as a quadratic Casmir operator, since the normal ordering regulates the summation over an infinite number of generators. By using Wicks Theorem (Eq.~\ref{eq:wick}), one can work through to find that

\begin{equation}
\contraction{}{J^a}{(z)}{T(}J^a(z)T(w) = (z-w)^{-2}J^a(w).
\label{eq:sugawara1}
\end{equation}

Transforming to $x$ space, as in Eq.~\ref{eq:exp} and defining 

\begin{equation}
\tilde{T}(z)= 4 \pi i \left( \frac{z}{L} \right)^2 T(z)
\label{eq:sugawara1}
\end{equation}

we can express our commutator relations as

\begin{equation}
\begin{split}
[\tilde{T}(x),\tilde{T}(y)] = \frac{c}{6\pi i}  \delta ''' (x-y) + 2 \tilde{T}(y) \delta' (x -y).
\end{split}
\label{eq:sugcom}
\end{equation}	

As with the current operator, we can expand $T(z)$ as a Laurent series

\begin{equation}
T(z) = \sum_{n \in \mathbb{Z}}L_n z^{-n-2},
\label{eq:sugawara2}
\end{equation}

and in terms of commutators, the $L_n$ terms become

\begin{equation}
[L_n,L_m] =  (n-m)L_{n+m} + \frac{n(n^2-1)}{12}c \delta_{n,-m}
\label{eq:sugawara3}
\end{equation}

By interpreting $c$ as a the eigenvalue of some central generating operator this commutator defines a \emph{Virasoro algebra} ($Vir$) which is an infinite dimensional Lie algebra given as a central extension of polynomial vector fields on $S^1$. The value $c$ is often called the \emph{conformal anomaly} or the \emph{conformal central charge}. 

	Mack and Luscher, in an unpublished paper in 1976, showed that the smallest positive value of $c$ which admits a unitary (i.e. information preserving) representation was $\frac{1}{2}$. It was not until eight years later, with the development of the Kac determinant, that Fuchs and Fegien showed that this for each $c$ which is expressible as $1 - 6 \frac{(p-p')^2}{pp'}$, with $p$ and $p'$ being coprime positive integers, that one can generate a finite set of fields that span an operator product algebra. These sorts of conformal theories are often denoted RCFTs, for rational conformal field theories. This subcategory of CFTs is particularly interesting since local observable fields are rational functions of the space-time coordinates, and therefore can be generalized to higher dimensional conformal field theories. \cite{todorov2001two}


\section{The Conformal Group}

A general conformal transformation, by definition, can be defined as an invertible mapping which leaves the metric tensor, $g_{\mu\nu}$, invariant up to some scale factor. We can fairly quickly identify the Poincare group (translations and rotations) as a necessary subgroup of the conformal group, but we also have two other symmetries, dilation and special conformal transformations defined as

\begin{equation}
x'^{\mu} = \frac{x^\mu - b^\mu x\cdot x}{1-2b \cdot x +b^2 x\cdot x}
\label{eq:sct}
\end{equation}

This transformation arises from the angle-preserving nature of conformal transformations. In general, we can express a transformation as $x'^\mu = x^\mu + \epsilon f^\mu(x)$, for some small $\epsilon$ and vector function $f$. Angle preservation follows naturally from $f$ satisfying the differential equation

\begin{equation}
d(\partial_\mu f_\nu + \partial_\nu f_\mu) - 2 \delta_{\mu\nu} \sum_{\rho =1}^{d} \partial_\rho f^\rho = 0
\label{eq:angles}
\end{equation}

with d being the dimension of the space. Eq.~\ref{eq:sct} falls naturally from solving this equation. The generators of these groups, say $N_a$ can be found from the transformation equations by looking at the infinitesimal transformation at a point of some field, $\Phi$,

\begin{equation}
\delta \Phi = \Phi ' (x) - \Phi (x) = -i N_a \Phi (x).
\label{eq:generator}
\end{equation}

Using this definition we get the differential realization for the generators of the conformal group, 

\begin{equation}
\begin{split}
P_\mu = -i \partial_\mu \\
L_{\mu\nu} = i(x_\mu \partial_\nu - x_\nu \partial_\mu)\\\
D = -i x^\mu \partial_\mu \\
K_\mu = -i(2x_\mu x^\nu \partial_\nu - x \cdot x \partial_\mu)
\end{split}
\label{eq:generators}
\end{equation}

consisting of translations, rotations, dilations and special conformal transformations (respectively). Notice that there is a problem with relating these transformations with $R^d$, as the special conformal transformations can map finite points to infinity. However, by compactifiying $R^d$ with this point, we can use the conformal group to cover the Lorentzian group $SO(d-2,2)$. This relation can be seen explicitly in terms of commutators. Notice that for the original generators we have the relations
\begin{equation}
\begin{split}
[D, P_\mu] =  i P_\mu, \\
[D, K_\mu] = -i K_\mu, \\
[K_\mu, P_\nu] =  2i(\eta_{\mu\nu}D-L_{\mu\nu}),\\
[K_\rho, L_{\mu\nu}] = i(\eta_{\rho \mu}K_{\nu} - \eta_{\rho\nu} K_{\mu}),\\
[P_\rho, L_{\mu\nu}] = i(\eta_{\rho \mu}P_{\nu} - \eta_{\rho \nu} P_{\mu}).
\end{split}
\label{eq:commutators}
\end{equation}
We can compress these into a more suggestive form by first introducing an antisymmetric generator $J_{ab}$ defined by
\begin{equation}
\begin{split}
J_{\mu\nu} = L_{\mu\nu},\\
J_{-1,0}= D,\\
J_{-1,\mu} = \frac{1}{2}(P_\mu - K_\mu),\\
J_{0,\mu} = \frac{1}{2}(P_\mu + K_\mu).\\
\end{split}
\label{eq:commutatorss}
\end{equation}
Using these relations we can construct a single commutator to represent the conformal group as
\begin{equation}
[J_{ab},J_{cd}] = i(\eta_{ad}J_{bc} + \eta_{bc}J_{ad}-\eta_{ac}J_{bd}-\eta_{bd}J_{ac}),
\label{eq:so(3)}
\end{equation}
which is also the $SO(d-2,2)$ commutator relation, proving the isogeny. However, this analysis is overly complex for the case of $d=2$. Eq.~\ref{eq:angles} simplifies drastically to just the Cauchy-Riemann equations
\begin{equation}
\partial_0 f_0 = \partial_1 f_1, \: \: \: \: \: \partial_0 f_1 = - \partial_1 f_0.
\label{eq:cr3}
\end{equation}

Combining these equations with complex notation ($\overline{z} ,z_0 = x_0 \mp i x_1$; $\overline{\partial}, \partial = \partial_0 \mp i \partial_1$; and $\overline{f},f = f_0 \mp i f_1$)
\begin{equation}
\overline{\partial} f(z,\overline{z}) = \partial \overline{f}(z,\overline{z}) = 0,
\label{eq:cr}
\end{equation}
so $f$ is analytic and will generate an infinite dimensional algebra. Using this constraint, we can construct the \emph{Witt algebra} in terms of the transformations defined by $f$ and $\tilde{f}$, with generators of the form $\hat{L}_n = -z^{n+1} \frac{d}{dx}$, and $\hat{T}(z)$ defined analogously to Eq.~\ref{eq:sugawara2}, we get the commutator brackets

\begin{equation}
[\hat{T}(x),\hat{T}(y)] = \delta ' (x - y) \hat{T}(y). 
\label{eq:cr}
\end{equation}

Via the \emph{Luscher-Mack theorem} one can add onto this commutator relation a conformal charge term, and thereby classify it as a Virasoro Algebra, since the Wightman axioms (uniqueness of vacuum state, locality of interactions, positivity of the metric, Lorentz invariance, and the spectrum condition) hold, the theory displays dilatation invariance, and there is a conserved symmetric energy-momentum tensor. 

	This energy-momentum tensor, $T_{\mu\nu} $, is supplied by  the Sugawara construction, which can be shown through explicit construction. Recall that an energy-momentum tensor is proportional to the variation of the action with respect to a world sheet metric, $\delta S / \delta g^{\mu\nu} = T_{\mu\nu}$, with the properties
\begin{equation}
\begin{split}
T_{\mu\nu} = T_{\nu\mu},\\
\sum_\mu \frac{\partial T_{\mu\nu}}{\partial x^\mu} = 0.
\end{split}
\label{eq:set}
\end{equation}
Dilatation invariance implies that the action be invariant under any arbitrary rescaling of the metric, $\delta g_{\mu \nu} = \epsilon g_{\mu \nu}$, so one can conclude that 
\begin{equation}
g_{\mu\nu} \frac{\delta S}{\delta g_{\mu \nu}} = 0 \implies  T_{\nu}^\nu = 0.
\label{eq:action}
\end{equation}
There remains only two non-zero terms
\begin{equation}
\begin{split}
T = T_{00} - T_{11} + 2i T_{01},\\
\tilde{T} = T_{00} - T_{11} - 2i T_{01}.
\end{split}
\label{eq:non-zero}
\end{equation}
Making the association of $T = T(z)$ and $\tilde{T} = \tilde{T}(\tilde{z})$ (the holomorphic and anti-holomorphic parts of Sugawara construction) provides a conformally invariant space with a energy-momentum tensor. Our complete algebra is therefore constructed by a direct sum of the Virasoro algebras associated with each term of the energy momentum tensor, and we get fields of the form $\phi(z,\tilde{z})$.

One can make the connection between the Sugawara construction and the original energy momentum tensor more concrete by considering the case of a two-dimensional free bosonic field. While most of machinery developed in section 3 and 4 is unnecessary to understand this special case since the algebra simplifies tremendously, it is still valuable to see the equivalence. In Euclidean space the Lagrangian takes the form
\begin{equation}
\mathcal{L} = \frac{1}{2} d^2x \partial_{\mu}\phi \partial_{\nu}\phi = dz d\tilde{z} \partial_z \phi \partial_{\tilde{z}} \phi.
\label{eq:Lagrangian}
\end{equation}
In this case, Eq.~\ref{eq:cr3} yields a decoupled relationship for the $\phi$ field as (note the symmetry of $z$ and $\tilde{z}$)
\begin{equation}
\phi(z,\tilde{z}) = \phi(z) + \tilde{\phi}(\tilde{z}),
\label{eq:fields}
\end{equation}
and current fields given in Eq.~\ref{eq:C-Mx} can be defined as
\begin{equation}
J^{a}(z) = \sum_{n \in \mathbb{N}} T^{a}_nz^{-n-1}=2i\sqrt{\pi}\partial_z \phi.
\label{eq:fields}
\end{equation}
It can be checked via Eq.~\ref{eq:integral} that the resulting $T^{a}_n$ terms follow the algebra structure given in Eq.~\ref{eq:constant}. Given this form for the affine current operator, we can construct an energy momentum tensor via Eq.~\ref{eq:sugawara},
\begin{equation}
\begin{split}
T(z)= \lambda \: \kappa^{ab}:J^a(z)J^b(z): = \lambda \delta^{ab} (-4\pi) : \partial_z \phi \partial_z \phi : \\
=  \lambda (-8\pi) \partial_z \phi \partial_z \phi
\end{split}
\label{eq:sugawaran}
\end{equation}
The choice of $\lambda$ is up to convention, although it is usually given in terms of the dual Coxeter number of the underlying affine Lie algebra. In this case it suffices to normalize the energy momentum tensor with $\lambda= (-16 \pi)^{-1}$ to give the traditional result, with a similar form for the $\tilde{z}$ component,
\begin{equation}
T(z)= \frac{1}{2}\partial_z \phi \partial_z \phi.
\label{eq:final_se}
\end{equation}
For the general system in order to properly define a conformal field theory completely in terms of the $\phi$ fields, it is required that a product of two fields satisfies the so-called \emph{bootstrap} condition, which was used implicitly in Eq.~\ref{eq:R_E1},
\begin{equation}
\phi_i(z,\tilde{z})\phi_j(w,\tilde{w}) = \sum_k d_{ij}^{k}(z,\tilde{z},w,\tilde{w}) \phi_k(w,\tilde{w}).
\label{eq:bootstrap}
\end{equation}
This is known as the \emph{operator product algebra}. This decomposition is extremely important when dealing with branching rules in conformal field theories and when using conformal blocks to expand four-point correlation functions in either chiral or conformal field theories.

\section*{Acknowledgements}

I would like to give thanks to Professor Francesco Iachello for his guidance and the opportunity to write this paper, and to Professor David Poland and Professor Igor Frenkel for sparing time to discuss conformal field theories with me. I would also like to thank the developers of TikZ for their excellent diagram creating scripts for \LaTeX.

\nocite{*}
\bibliographystyle{plain}
\bibliography{cft}



\end{document}